%% file: main.tex
\documentclass[5p,sort&compress]{elsarticle}
\usepackage{hyperref}
\usepackage{lineno}
\usepackage{natbib}
\usepackage[english]{babel}
\usepackage{textcomp}
\modulolinenumbers[5]
\usepackage{amssymb}
\usepackage{amsmath} 
\usepackage{framed} 
\usepackage{cancel}
\usepackage{booktabs}

\usepackage{multicol} 
\usepackage{nomencl} 
\usepackage{etoolbox}

\include{_preamble}

\journal{Energy \& Buildings}

\begin{document}

\begin{frontmatter}

\title{Bottom-up energy supply optimization of a national building stock}



%


\author[addFZJ,addLBL]{Leander~Kotzur\corref{corLK}}
\author[addFZJ]{Peter~Markewitz}
\author[addFZJ]{Martin~Robinius}
\author[addLBL]{Gon{\c c}alo~Cardoso}
\author[addFZJ]{Peter~Stenzel}
\author[addLBL]{Miguel~Heleno}
\author[addFZJ,addCFC]{Detlef~Stolten}

\cortext[corLK]{Corresponding author. Email: l.kotzur@fz-juelich.de}
\address[addFZJ]{ Institute of Electrochemical Process Engineering (IEK-3), Forschungszentrum J\"ulich GmbH, Wilhelm-Johnen-Str., 52428 J\"ulich, Germany}
\address[addLBL]{Lawrence Berkeley National Laboratory, University of California, 1 Cyclotron Rd, Berkeley, CA, USA}
\address[addCFC]{Chair for Fuel Cells, RWTH Aachen University, c/o Institute of Electrochemical Process Engineering (IEK-3), Forschungszentrum J\"ulich GmbH, Wilhelm-Johnen-Str., 52428 J\"ulich, Germany}

\begin{abstract}
The installation and operation distributed energy resources (DER) and the electrification of the heat supply significantly changes the interaction of the residential building stock with the grid infrastructure. Evaluating the mass deployment of DER at the national level would require analyzing millions of individual buildings, entailing significant computational burden. 

To overcome this, this work proposes a novel bottom-up model that consists of an aggregation algorithm to create a spatially distributed set of typical residential buildings from census data. Each typical building is then optimized with a Mixed-Integer Linear Program to derive its cost optimal technology adoption and operation, determining its changing grid load in future scenarios.

The model is validated for Germany, with 200 typical buildings considered to sufficiently represent the diversity of the residential building stock. In a future scenario for 2050, photovoltaic and heat pumps are predicted to be the most economically and ecologically robust supply solutions for the different building types. Nevertheless, their electricity generation and demand temporally do not match, resulting in a doubling of the peak electricity grid load in the rural areas during the winter. The urban areas can compensate this with efficient co-generation units, which are not cost-efficient in the rural areas.
\end{abstract}


\end{frontmatter}

\makenomenclature
\setlength{\nomitemsep}{-0.05in} 
\renewcommand*\nompreamble{\begin{multicols}{2}}
\renewcommand*\nompostamble{\end{multicols}}

\input{_nomenclature}

\begin{table*}[!t]   
\begin{framed}
\printnomenclature
\end{framed}
\end{table*}


%
\input{01_Introduction}

\input{02_Method}

\input{03_Validation}

\input{04_Results}

\input{05_Conclusion}

\section*{Acknowledgments}
\label{s:Acknow}
This work was supported by the Helmholtz Association under the Joint Initiative  \href{https://www.helmholtz.de/forschung/energie/energie_system_2050/}{\textit{Energy~System~2050 $-$ A Contribution of the Research Field Energy}}. Parts of this article are derived from the doctoral thesis \href{http://hdl.handle.net/2128/21115}{\textit{Future Grid Load of the Residential Building Sector}}.

\appendix
\label{sec:Appendix}
\input{00_Appendix}

\bibliography{mybibfile}

\end{document}

%% file: _preamble.tex
\newcommand{\myHereFigure}[4][image]%
{%
\begin{figure}[h]
	\begin{center}%
		\includegraphics{#2}
		\caption[#4]{#3}%
		\label{#1}%
	\end{center}%
\end{figure}%
}%

\newcommand{\myHereWidthFigure}[4][image]%
{%
\begin{figure}[h]%
	\begin{center}%
		\includegraphics[width=0.8\columnwidth]{#2}%
		\caption[#4]{#3}
		\label{#1}%
	\end{center}%
\end{figure}%
}%

\newcommand{\myTwoColFigure}[4][image]%
{%
\begin{figure*}[tbh]%
\centering
		\includegraphics[width=1.92\columnwidth]{#2}%
		\caption[#4]{#3}
		\label{#1}%
\end{figure*}%
}%

\newcommand{\myTable}[4]%
{%
\begin{table}[h]%
	\caption[#4]{#2}%
	\begin{center}%
			#1%
		\label{#3}%
	\end{center}%
\end{table}%
}%

\newcommand{\myScaledTable}[4]%
{%
\begin{table}[h]%
	\caption[#4]{#2}%
	\begin{center}%
		\resizebox{0.95\columnwidth}{!}{
			#1%
		}
		\label{#3}%
	\end{center}%
\end{table}%
}%

\addto\captionsenglish{}

%% file: _nomenclature.tex
\nomenclature[A]{LB}{Lower Bound}{}{} 
\nomenclature[A]{UB}{Upper Bound}{}{}
\nomenclature[A]{CRF}{Capital Recovery Factor}{}{}
\nomenclature[A]{WACC}{Weighted Average Cost of Capital}{}{}
\nomenclature[A]{SFH}{Single-Family House}{}{}
\nomenclature[A]{MFH}{Multi-Family House}{}{}
\nomenclature[A]{CAPEX}{CAPital EXpenditure}{}{}
\nomenclature[A]{OPEX}{OPerational EXpenditure}{}{}
\nomenclature[A]{GHG}{Greenhouse Gas}{}{}
\nomenclature[A]{CHP}{Combined Heat and Power}{}{}
\nomenclature[A]{DER}{Distributed Energy Resources}{}{}
\nomenclature[A]{RMSE}{Root Mean Squared Error}{}{}
\nomenclature[A]{QIP}{Quadratic Integer Program}{}{}

\nomenclature[B]{$D$}{Scaling of a technology [kW(h)]}{}{}
\nomenclature[B]{$\dot{E}$}{Energy flow [kW]}{}{}
\nomenclature[B]{$SOC$}{State of charge [kWh]}{}{}
\nomenclature[B]{$\delta$}{Existence of a technology [-]}{}{}
\nomenclature[B]{$\gamma$}{Greenhouse gas footprint [kg/kWh]}{}{}

\nomenclature[B]{$\Delta t$}{Duration of a single time step [h]}{}{}
\nomenclature[B]{$\tau$}{Economic lifetime [a]}{}{}
\nomenclature[B]{$\eta$}{Efficiency or quality grade [-]}{}{}

\nomenclature[C]{$L$}{Set of technology connections}{}{} 
\nomenclature[C]{$d$}{Index of the a general device or component}{}{}
\nomenclature[C]{$t$}{Index of the discrete time step}{}{}
\nomenclature[C]{$q$}{Index of the Source/Sink class}{}{}
\nomenclature[C]{$n$}{Index of the Collector class}{}{}
\nomenclature[C]{$f$}{Index of the Transformer class}{}{}
\nomenclature[C]{$s$}{Index of the Storage class}{}{}
\nomenclature[C]{$\epsilon$}{Energy or commodity type}{}{}

%% file: 01_Introduction.tex
\section{Introduction}
\label{sec:Introduction}

\subsection{Background}

The residential building sector is responsible for 17\% of worldwide CO$_2$ emissions \cite{IEA2015}. In Germany, it was the source of 10\% of the total Greenhouse Gases (GHG) through fossil fuel combustion in the year 2015. Moreover, it was responsible for 12\% of total emissions due to the GHG footprint of its energy imports \citep{AGEB2017,rwi2017,BMWi2016}. These emissions must be cut in order to reach the overall goal of net zero GHG emissions in the second half of this century \citep{UN2015} with the goal of minimizing the impact of anthropogenic climate change \citep{Solomon2009}. Therefore, the European Union introduced the concept of "Zero Energy Buildings" (ZEB) in the context of its energy performance of buildings directive \citep{EU2010,EU2012} with the goal of deploying GHG-neutral buildings that compensate for their emissions by exporting on-site generated renewable energy \citep{REHVA2011,Marszal2011}.

While the objectives are clear, the pathway to GHG-neutral building stock is uncertain and the integration of new technological solutions unsettles utilities \citep{OCallaghan2014,Agnew2015} as well as governments \citep{Rickerson2014}. Therefore, analyses are needed that predict technological development and their system integration: Many works for the building sector \citep{McKenna2013,BMWi2015,Diefenbach2016,Beuth2017,BMWi2018} solely focus on GHG reduction strategies for heat demand. They conclude that significant energy saving potentials can be accessed by increased refurbishment rates and that residual heat can be supplied with renewable energy. 

Nevertheless, in relation the heat demand of the building sector can no longer be regarded as being more isolated from the other energy system: Heat pumps are seen as a key option to efficiently provide space heat \citep{IWES2015,UBA2017}, while Combined Heat and Power (CHP) generation allows an efficient usage of chemical energy carriers while providing flexibility to the grid \citep{Lund2012}. Furthermore, a trend towards an increased self-supply of residential buildings is apparent with the rapidly falling prices of photovoltaics \citep{ISE2015} and batteries \citep{Nykvist2015} constituting \textit{grid parity} \citep{Breyer2013}, meaning that the levelized cost of self-generated electricity is below the retail electricity grid price.

Both trends, i.e., the changing heat supply and the increasing self-sufficiency of the buildings, will significantly change the future grid demand and challenge the feasibility of current electric grid design. Therefore, new analyses are required that consider the adoption and operation of new supply technologies and efficiency measures, predicting the spatially- and temporally-varying impacts on the grid infrastructure.

\subsection{Related works}

Various works have already perforemed top-down analyses of the load change due to single Distributed Energy Resources (DER), like photovoltaic \citep{Mainzer2014}, heat pumps and battery electric vehicles \citep{ISI2016}, or Combined Heat and Power (CHP) \citep{Beer2012}. \citet{Seljom2017} analyze the impact of photovoltaics deployed in ZEB in for the future Scandinavien energy system. Nevertheless, as demand, flexibility options and generation are closely connected to the building supply systems, these technologies can not be evaluated independently and must be get holistically regarded and modeled as system entities. This can primarily be done by means of detailed bottom-up models that simultaneously consider investment decisions and the operation of DER as well as efficiency measures and demand side management. 


\subsubsection{Building optimization}

Thereby, the models must account for cost optimality, as the main motivations of building owners to adopt different supply technologies are savings or earnings emanating from their installation \citep{Balcombe2014}. This also applies for the efficiency measures or energy retrofits, where the need of replacement or financial profitability are the main activators for the adoption \citep{Achtnicht2014}. 

Therefore, many different optimization models have been proposed for determining the cost optimal investment decisions and operation of building supply systems: either as Linear Programs (LP) \citep{Milan2012, Lauinger2016} with the advantage of good computational tractability  but the disadvantage of not being able to account for economies of scale; or as a Mixed Integer Linear Program (MILP) model \citep{Ashouri2013,Mehleri2013,Fazlollahi2012,Harb2015,Lindberg2016,Cardoso2017,Schutz2017a}. Furthermore, two-level approaches that determine at least a part of investment decisions with a meta-heuristic solver and operation with a simulation or optimization are popular \citep{Hamdy2013,Evins2015,Stadler2016,Han2016,Streblow2017,Wu2017}. The last approach can account for very detailed physical models but no global optimal solution is guaranteed. Some of the models even include the investment decision into efficiency measures by changing the buildings envelope \citep{Stadler2014,Evins2015,Wu2017,Schutz2017}. 

All models would enable the analysis of the impact of technology adoption and operation on the local infrastructures: for instance, \citet{Lindberg2016} apply a MILP to design the supply system of a Multi-Family-House (MFH) and analyze the resulting electricity grid load for cost-optimal system operation under current German regulations. \citet{Schutz2017a} use a model to evaluate the optimal technology adoption with currently considered incentives and market conditions for the case of three reference buildings.

\subsubsection{Archetype buildings}

Although these analyses and models already provide many insights for the application cases, a further generalization would be required to upscale these results to an aggregated nationwide perspective. Furthermore, the spatial variation due to  regionally differing building topologies would be required to integrate the results to grid models. Therefore, a set of representative buildings is required that characterizes the spatial diversity of building stock and that can be used for the previously described building models. 

In general, such typical buildings are often referred to as \textit{archetype buildings} and are commonly used for modeling GHG reduction strategies in the building sector \citep{Kavgic2010}, as described in the Energy Performance of Buildings Directive \citep{EU2010,EU2012} of the \textit{European Union}.

In this context, \citet{Corgnati2013} introduced different pathways to determine representative reference buildings for the analysis of cost optimal refurbishment measures, but conclude that in reality most often a mixture is used due to the different available data for different buildings stocks.

\citet{Mata2014} proposed an analytical methodology to aggregate archetype building stocks based on publicly available data. The steps include a \textit{segmentation} based on categories such as construction year, a technical \textit{characterization} such as the thermal transmittance as input values for energy performance models, as well as a \textit{quantification} to scale the buildings up to a nationwide level. The methodology is applied to France, Germany, Spain, and the UK, with the resulting final energy demand showing a deviation of less than -4\% to + 2\% to the aggregated statistical values.

Various nation-specific works exist to quantify the energy consumption of the building sector with the help of archetype buildings: the German residential building stock is described by a framework developed by the \textit{Institute für Wohnen und Umwelt} (IWU) \citep{IWU2005,IWU2010}. This schema categorizes the stock into classes that differ by construction year and building size that are represented with detailed technical parameters. This stock description has been extended to other European countries in the framework of the \textit{EPISCOPE} project \citep{EPISCOPE2016}. The US \textit{Department of Energy} (DOE) introduced archetype buildings for the residential sector \citep{Hendron2010} and the service sector \citep{Torcellini2008} in the USA, referred to as benchmark or prototype buildings. An advantage over the European database \citep{EPISCOPE2016} is the fact that additional time series data are provided for the different building demands, including electricity, hot water, cooling or heating demand for typical days in different climate zones.

The aggregation to archetype buildings is also widespread in the context of urban energy models: \citet{DallO2012} present a work flow to derive archetype buildings with a combination of statistical data and a survey applied to sample buildings. Meanwhile. \citet{Cerezo2015} and \citet{Sokol2017} introduce methods to estimate unknown attributes of the proposed archetype buildings, such as comfort temperature levels, based on a probability distribution. This approach can make use of measured energy data in different buildings, such as annual or monthly gas demand, and fits the uncertain attributes to it. Moreover, \citet{Fazlollahi2014} and \citet{Fonseca2015} use k-means clustering methods to group similar buildings in urban areas by using the location of the buildings and spatially resolved statistics. The advantage of clustering is that the simulation models or optimization models can be applied to the zones instead of the single buildings, which reduces the number of variables and the computational load of the related models. 




\subsection{Own approach and structure}

In summary, many works exist that consider detailed integrated building optimization models to determine the cost optimal technology adoption and operation. There also exist many different approaches for the aggregation and segmentation of archetype buildings. Nevertheless, to the knowledge of the authors the combination of both for the purpose of analyzing the spatially and temporally changing demand of the infrastructure, i.e. gas grid and electricity grid, does not exist. 

\subsubsection{}

Therefore, this work proposes a two stage framework to 
\begin{enumerate}
    \item \textbf{aggregate} a spatially resolved building stock with a limited number of \textbf{archetype buildings} which are described by a set of attribtues related to their energy supply and demand
    \item and \textbf{to optimize} those different buildings in parallel, considering a superstructure of \textbf{supply technologies and potential efficiency measures}.
\end{enumerate}
Thereby, different regulatory regimes can be considered, resulting in different future technology installations and operations, and grid demands. The spatial assignment of the archetype buildings allows then for the local evaluation of the changing energy demand. The general idea is visualized Figure \ref{fig:structure_idea}.

\myHereWidthFigure[fig:structure_idea]{images//introduction_Structure_3}{Structure of the bottom-up model to optimize the a spatially resolved building stock and determine the changing infrastructure usage.}{Building archetype aggregation algorithm}%

With the help of this modeling approach, the potential for self-sufficient energy supply systems in residential buildings can be efficiently evaluated and its large-scale techno-economic impact on the grid demand can be derived. To do so, Section~\ref{sec:Method} introduces the used aggregation tool, the used data sets and the parallel building optimization for this model. The approach is applied for a reference scenario in Section~\ref{sec:Validation} and validated to the available aggregated energy demand data in Germany. Thereby, a trade-off is made regarding the number of archetype buildings to describe the diversity of building stock on one hand, but limiting the computational by the number of Mixed-Integer Linear Programs to solve. In order to derive the changing load, a future supply scenario is introduced in Section~\ref{sec:res_mincost} for the year 2050. Section \ref{sec:Conclusion} critically recaps the work and draws the main conclusions.

%% file: 02_Method.tex
\section{Methods and data}
\label{sec:Method}

The model consists of two main components: 
\begin{enumerate}
	\item The building optimization model, described in Section~\ref{sec:singlebuildingopt}, represents the decision making of the building owner regarding the design and operation of the energy supply and energy demand.
	\item The aggregation and distribution of archetype buildings, introduced in Section~\ref{sec:buildingaggregation}, makes the building optimization results generalizable to the perspective of regulators and grid operators on a nationwide scale.
\end{enumerate}

\subsection{Building optimization}
\label{sec:singlebuildingopt}

It is assumed that building owners are the decision makers and they consider the energetic supply of the building from the perspective of a single economic entity.
Therefore, the goal of the building optimization is to have a holistic perspective of single residential buildings and is able to consider synergies between different solutions, e.g. demand-side measures are simultaneously considered with supply-side measures, or the operation of the heating system is optimized together with the operation of the electricity system. 

\subsubsection{Creation of demand and supply time series}

In order to derive the cost optimal supply system for each individual building, first the temporally varying energy demands of electric devices, hot water and thermal comfort, as well as the varying performance of renewable supply technologies are derived as follows:

The occupancy behavior and inherited electricity load of the appliances are created with the help of the CREST demand model \citep{McKenna2016,Richardson2010,Richardson2009,Richardson2008}. Further, the demand for hot water is separated from it. The advantage of this model is that it sufficiently incorporates the high variance of single residential load profiles, as well as the stochastic smoothing for the case that an agglomeration of households is considered. A validation of the model in the context of German residential electricity demand is performed in \citet{Kotzur2018} and exhibits sufficient accuracy.

The heat load is considered with the 5R1C-model from EN ISO 13799 \citep{EN2008}, which was implemented into a MILP by \citet{Schutz2017}. The physical building properties, such as heat transfer coefficients for different construction years are taken from \citet{IWU2010}. This is able to account for the thermal building mass for a flexible supply system operation. Furthermore, potential refurbishment measures are part of the solution space, such as the addition of wall or roof insulation, the replacement of windows, or the integration of smart thermostats. The configuration of the buildings is introduced in detail in \citet{Kotzur2018}.

The time series for PV and solar thermal are created with the PV-Lib \citep{Andrews2014}. The weather data is derived from the DIN EN 12831 \citep{DIN2014} by finding the closest location listed. Therefrom, the minimal design temperature is derived as well as the test reference region of the \textit{Deutsche Wetter Dienst (DWD)} \citep{DWD2012}. Alternatively, the weather from the COSMO rea-6 reanalysis data set \citep{Bollmeyer2015} is used for real weather years for validation purposes. 

The whole initialization of the building specific time series are published in the open source python package \textit{tsib - Time Series Initializaion for Buildings} (\url{https://github.com/FZJ-IEK3-VSA/tsib}).

\subsubsection{Optimizing structure, scale and operation}

The building optimization is based on a typical Mixed-Integer Linear Program (MILP) with the objective of minimizing the annual energy cost of a single building as proposed in the vast literature. The operation and design of the supply system is modeled with the object-oriented system modeling framework FINE \citep{Welder2018,Kotzur2018,Kotzur2018a}. The binary variables are considered to sufficiently incorporate the economy of scale of the technologies. The operation is modeled in a fully linear and continuous manner for 8760 hours in a representative year. 

All in all, the combinatorial consideration of demand-side and supply-side measures respecting the full operational variety yields a complex mathematical program that is computationally demanding. In order to keep the program tractable for many different building types and scenarios, the annual time series of weather, occupancy behavior, and appliance load are aggregated to twelve typical days with a hierarchical aggregation \citep{Kotzur2018a,Kotzur2018aa}. The days with the smallest temperature and highest electricity load are added as extreme days. Based on these, the optimal choice of the supply technologies and refurbishment measures is determined with the MILP. The binary decision variables are then fixed and a validation and scaling optimization is then performed for the full annual time series \citep{Kannengieer2019}, similar to that in \citet{Bahl2017}. 

The detailed model description and their independent validation can be found in \citet{Kotzur2018}.

\subsection{Aggregation of archetype buildings}
\label{sec:buildingaggregation}

In order to determine the different types of buildings the number of their occurance, this section introduces a new aggregation method to derive spatially-distributed archetype buildings. Those are used to scale the results of building optimizations to a spatially-resolved nationwide perspective. Therefore, Section~\ref{sec:agg_attributes} discusses the relevant attributes to describe the energy performance of a building. The aggregation algorithm itself is sketched in Section~\ref{sec:agg_algorithm} and its application is illustrated in Section~\ref{sec:agg_app}.

\subsubsection{Relevant building attributes}
\label{sec:agg_attributes}

In general, four categories of building attributes are emphasized in the literature, while the concrete nomenclature varies \citep{Corgnati2013,Mata2014}: The \textit{Form} describes the physical exterior shape of the building, including orientation, wall area, window area and roof areas. The \textit{Envelope} characterizes the physical properties of the materials used in the building. The technologies installed in the buildings to satisfy thermal comfort and other demands are grouped into the category of \textit{System}. \textit{Operation}, in turn, summarizes all extrinsic conditions determining the system operation, such as the local weather or occupancy behavior.

Aside from the attributes describing the current energy performance, future energy supply is also of interest, where the category of \textit{Adoption} summarizes all attributes referring to the investment capabilities and investment behavior of the building owner, such as a potential interest rate. For instance, the model described here considers the cost-optimal technology adoption of the different buildings.

The categories and their aggregation define a general framework to segment buildings, but the required attributes depend on the model and data availability, e.g., the envelope could be described by materials with exact heat conductivities and thicknesses, only heat transfer coefficients, or by the construction year of the building from which these values are derived.

The attributes considered for the aggregation procedure of this work are oriented towards the model introduced in Section~\ref{sec:singlebuildingopt} and the data provided by the census \citep{Statistisches2011}. Figure \ref{fig:CensusDist} shows the aggregated Census data for Germany.  The total number of buildings is predominated by over 12.3 million Single-Family Houses (SFH) and 6.3 multi-family houses, while buildings with more than 12 apartments have, at 0.21 million, a small share at the total number. The majority of the SFH are detached, constituting an overall small proportion of terraced and semi-detached buildings. 23.2 million of the 40.5 million apartments are rented, while one- and two-person households dominate with together 27.1 million households. These also constitute the peak of apartment sizes, with compact living areas of 59 to 79 m$^2$ per household, while larger single-family houses are spread over a larger grouping.

\myHereWidthFigure[fig:CensusDist]{images//CensusDist}{Aggregated attribute distribution of the German residential building stock based on the Census \citep{Statistisches2011}.}{Aggregated census attribute distribution}%

All of these distributions are also available on an absolute scale for the municipalities or a 100m grid in Germany and state the data basis for the considered archetype aggregation, as shown in Figure \ref{fig:aggregation_attributes}. The introduction of additional values is discussed in detail in \citet{Kotzur2018}.

\myHereWidthFigure[fig:aggregation_attributes]{images//aggregation_parameters}{Structure of the considered attributes that relevant for the building energy supply.}{Aggregated building attributes}%

\subsubsection{Building aggregation algorithm}
\label{sec:agg_algorithm}

These attribute distributions must be aggregated to a limited set of archetype buildings to evaluate them energetically. 

In general, the attributes belonging to the different categories are primarily published as aggregated distributions for different administrative boundaries that define the spatial resolution. Nevertheless, the real building instances and their values are unknown wherefore cross-combinations of attributes are not reproducible, e.g., how many terraced buildings have a certain living area. 

Thereby, two challenges arise: first, the buildings are described by a mixture of categorical and continuous attributes. Approaches exist dealing with this type of aggregation class, such as the mixture of \textit{k-means} and \textit{k-modes} clustering, referred to as \textit{k-prototypes} \citep{Huang1998}. Nevertheless, these would rely on a data set consisting of real building instances that should be clustered and can then be represented, e.g., by its medoid. This does not apply to the attributes distributions, and so a new aggregation methodology is required that is introduced in \citet{Kotzur2018}.

Thereby, a greedy algorithm is introduced with the goal of determining a locally optimal set of archetype buildings. It is inspired by the concept of an expectation-maximization algorithm, where Lloyd's k-means clustering algorithm \citep{Lloyd1982,Jain2010} and the k-prototypes algorithm \citep{Huang1998} belong as well. 

\myHereWidthFigure[fig:aggregation_algorithm]{images//aggregation_algorithm}{Flow chart of the developed algorithm to determine a spatially distributed archetype building stock.}{Building archetype aggregation algorithm}%

The idea is to describe the assignment of the archetype buildings to the different nodes or municipalities as the \textit{expectation} step, with the objective of getting a representation of the attribute distributions by the most likely archetype buildings into every municipality. Nevertheless, the attributes of the archetype buildings themselves are unknown, and so their estimation is defined as the \textit{maximization} step, illustrated in Figure \ref{fig:aggregation_algorithm}. This results in two optimization problems that are iteratively solved. 

The result is a set of archetype buildings and a matrix that defines the representation of every municipality by the number of certain archetype buildings.

\subsubsection{Application and illustration of the results}
\label{sec:agg_app}

The initial guess of the archetype building attributes, the start solutions for the algorithm, are derived from the current state of the art archetype buildings for Germany \citep{IWU2010}, while missing parameters are randomly generated, e.g. the number of persons living in an apartment. 

In the ollowing, the algorithm is applied once to different predefined numbers of archetype buildings in order to determine how many of these are required for a sufficient representation of the German building stock.

The quality of the resulting representation of the different attributes for different numbers of archetype buildings is illustrated in Figure \ref{fig:ParamFit_bdgs}. This is defined as the cumulative deviation of the representation of an attribute expression $m$ for every region in proportion to the total attribute manifestations for the whole of Germany:
\begin{equation}
f(p,m) = \frac{ \sum_{n \in N}  d_{n,m,p} - \left| d_{n,m,p} - \sum_{b \in B}{\beta_{b,n} \delta_{b,p,m}} \right| }{\sum_{n \in N} d_{n,m,p}  } \quad \forall \quad  p, m 
\end{equation}

\myHereWidthFigure[fig:ParamFit_bdgs]{images//ParamFit_bdgs}{Relative fit of all categorical attribute expressions for different numbers of archetype buildings in Germany.}{Attribute fit archetype buildings}%

The figure shows that for some of the attribute expressions, a small number of archetype buildings is able to represent them sufficiently, for instance single-family houses with a single apartment or the energy supplied by gas boilers. These are attribute expressions that often occur in the original data set. Therefore, they are first represented by the archetype buildings to reduce the overall error.
Nevertheless, attributes such as a CHP, heat pump supply, or apartments with a living area smaller than 39 m$^2$ rarely occur in the Census data set. Therefore, the algorithm has a secondary priority to represent them and focuses instead on building attributes that exist more often, e.g., no archetype building was created with a heat pump supply for 5, 25 and 50 archetype buildings because the overall share of heat pump supply in Germany is below 2 \%. Thus, it would not be efficient to sacrifice an additional archetype building to represent it.

In general, a fit below 100 \% does not imply that the expression is highly under-represented on the aggregated level: While an overestimation of an attribute in one region and an underestimation in the other regions constitute a reduced fit, they could add up and compensate each other on an aggregated nationwide level, which is further elaborated in \citet{Kotzur2018}. 

\begin{figure}[h]%
	\begin{center}%
		\includegraphics[width=0.5\columnwidth]{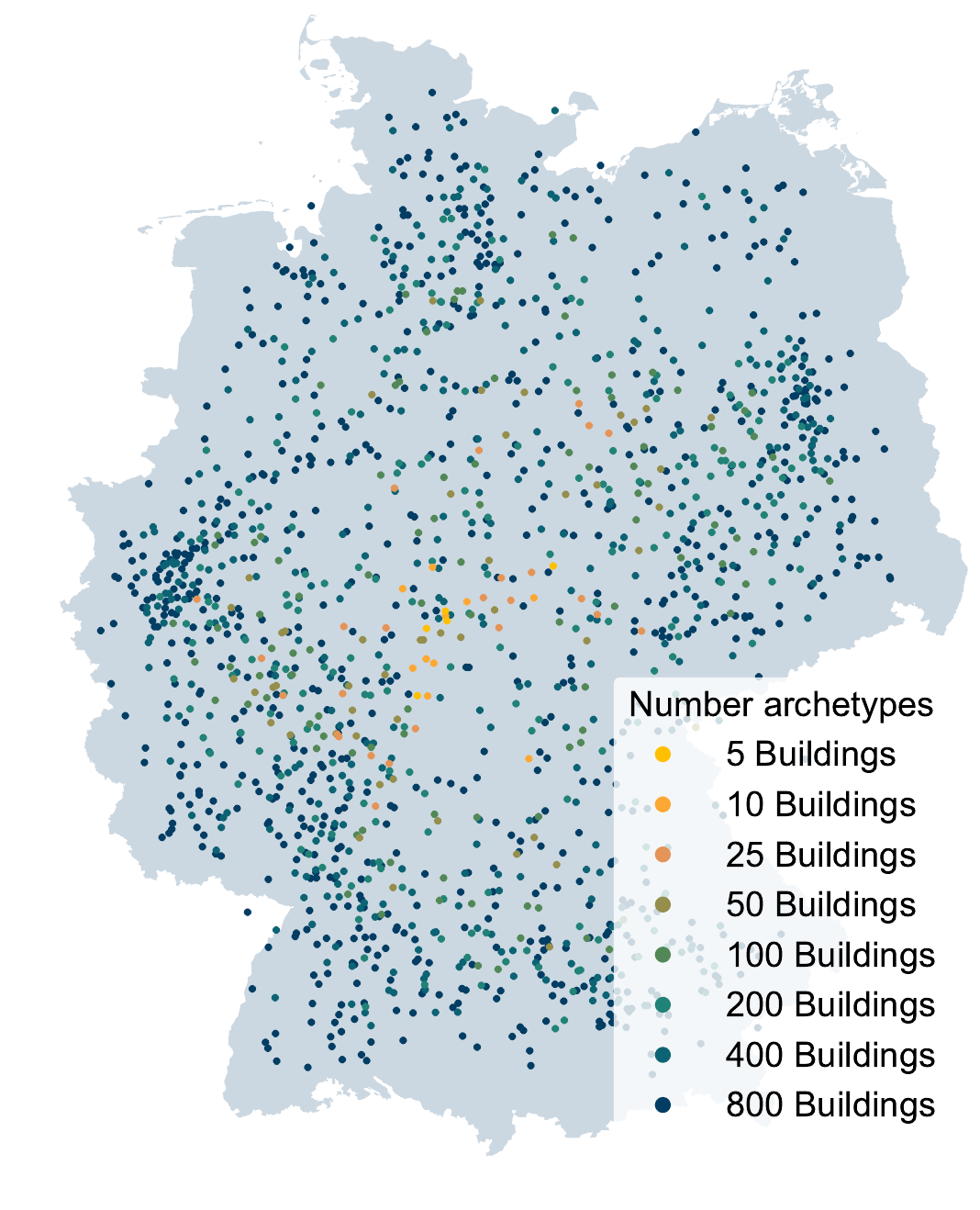}%
		\caption[Location archetype buildings]{Geographical location of different numbers of archetype buildings in the final iteration step.}
		\label{fig:buildingLocation}%
	\end{center}%
\end{figure}

The fitting of the continuous attributes, the latitude and longitude, is qualitatively illustrated in Figure \ref{fig:buildingLocation} with their exact geographical placement in Germany. For the case of 5 to 25 archetype buildings, all buildings are primarily located in the center of the country. The reason is that building archetypes are mainly used to represent the diversity of categorical attribute combinations that are spatially distributed across Germany. E.g., a single-family house from 1960 with a four-person household and gas boiler supply manifests as an archetype building that represents this building type in the north as well as in the south. For higher numbers of archetype buildings from 100 to 800, the geo-spatial location of the archetype buildings is spreading, as similar categorical building types can be instantiated multiple times. For the case of 800 archetype buildings, it is even observable that urban areas are represented by more archetype buildings than rural areas.

The representation of municipalities is further illustrated in Figure~\ref{fig:buildingAppearance}, which introduces the local assignment to the different municipalities of certain example archetype buildings, selected from the set of 800. The locations of the archetype buildings are the centers of the buildings they represent in the different municipalities. As these representations are spatially spread over different municipalities, the archetypes are not placed at municipalities at the border. 

Furthermore, it is recognizable that the areas and amounts that are represented differ between the archetypes: While an archetype building supplied with heat pumps must represent buildings over a large area, archetype single-family houses supplied by gas boilers have a definite local assignment area. The reason for this is that more archetype buildings with gas boilers are selected, as more buildings with gas boilers also exist in reality. Therefore, the algorithm chooses a higher spatial separation for them to minimize the overall error, while accepting a higher geospatial estimation error for the few buildings with heat pumps. 

\myHereWidthFigure[fig:buildingAppearance]{images//buildingAppearance_800}{Location of the most northern (blue) and most southern (green) single-family house archetype with heat pump supply (left) and gas boiler supply (right) of the set of 800 archetype buildings, and their assignment to the different municipalities.}{Local assignment of archetype buildings}%


%% file: 03_Validation.tex
\section{Validation of the method}
\label{sec:Validation}

For the validation of the two-staged methodology, the different numbers of aggregated archetype buildings are independently optimized for the status quo and then multiplied with their appearance in Germany. The choice of the technologies is predefined by the archetype definition \citep{Kotzur2018}, but the technology scale and operation are optimized such that the building-specific energy demands are met. It defines the \textsl{Reference}, or status quo, of the residential energy supply and validates the model to national energy demand statistics. The techno-economic assumptions are introduced in \ref{sec:res_assump}. The different sets of archetype buildings together with their spatial distribution can also be found in the \textit{tsib} (\url{https://github.com/FZJ-IEK3-VSA/tsib}).




\subsection{Impact of the number of archetype buildings}
\label{sec:res_validation_nobdgs}

The impact of choosing different number of archetype buildings to the resulting final energy demands, aggregated to different energy carriers, is illustrated in Figure~\ref{fig:EnergyDemandBdgVar}. They are validated against the final energy demand provided by \citet{AGEB2017}.

\myHereWidthFigure[fig:EnergyDemandBdgVar]{images//EnergyDemandBdgVar}{Final residential energy demand predicted for different numbers of archetype buildings.}{Final energy demand of archetype buildings}%

The dominant energy carriers for the residential sector are gas, oil and electricity with 268, 162 and 136~TWh/a per year \citep{AGEB2017}. The demand for renewable energy or district heating is secondary with 84 and 51~TWh/a per year. As can be seen in Figure~\ref{fig:EnergyDemandBdgVar}, the model is able to roughly predict with five to ten archetype buildings the demand of the three dominant energy carriers, but the appearance of minor energy supply carriers is not sufficiently included. This improves with an increasing number of archetype buildings while the best fit can be achieved with 800 buildings. Nevertheless, the resulting demands of 64.8~TWh/a for renewables and 44~TWh/a for district heating are still underestimated. This deviation is constituted by the aggregation, which tries to capture the most frequently appearing archetype buildings and neglects rarely occurring building types. 
Nevertheless, these missing energy demands for renewables and district heating are compensated by gas and oil demands, which are slightly overestimated with 286 and 165~TWh/a. This compensation effect already appears for 25 archetype buildings, where all cases between 25 and 800 archetype buildings predict the total final energy demand in a similar magnitude as the \citet{AGEB2017}. Above 200 archetype buildings, the share of the different energy carriers also aligns well with the structure of the \citet{AGEB2017}. 

The prediction with 50 archetype buildings overestimates the demand for oil by 27.4\% and underestimates the demand for gas by 17.1\%, while 100 archetype buildings on the contrary overestimate the gas demand by 25.5\% and underestimate the oil demand by 16.1\%. This switch shows a drawback of the aggregation: Some archetype buildings appear often and therefore have a high impact on the overall energy load. If the majority of the buildings supplied, e.g., with gas boilers have a construction year before 1960 while the more modern buildings are supplied with oil, an overestimation of the gas demand and an underestimation of the oil demand results, although the absolute number of the different boiler types is well represented. Nevertheless, this effect is reduced with an increasing number of archetype buildings, as single archetypes represent attribute distributions on a more granular level. In consequence, the spatial differences, e.g., of construction years, are better fitted and intrinsic correlations of the input data are represented with higher accuracy.

This work assumes 200 archetype buildings as a sufficient trade-off between accuracy and computational load, since they already capture the main diversity of the energy carriers and the statistical balancing effects between the buildings.

\subsection{Impact of different weather years}
\label{sec:res_validation_years}

The impact of different weather years from 2010 to 2015 on the energy demand of the buildings is illustrated in Figure~\ref{fig:EnergyDemandYearVar} for 200 archetype buildings and validated again to the final energy demand values provided by \citet{AGEB2017}. According to \citet{AGEB2017}, the total residential energy demand varies from 743~TWh in 2010 at a maximum to 608~TWh for 2014 as a minimum.

\myHereWidthFigure[fig:EnergyDemandYearVar]{images//EnergyDemandYearVar}{Final energy demand for different weather years predicted with 200 archetype buildings and compared to the values reported by the \citet{AGEB2017}.  }{Final energy demand for different weather years}%

For all different weather years, the systematic overestimation of gas demand and underestimation of district heating demand is observed, as already discussed in the previous section. Nevertheless, the relative deviation differs between the years. While the total final energy demand fits well for 2010 with an underestimation of below 2\%, the deviation increases in the year 2011 up to 7.4\%. It reduces again to 2.8\% in the year 2013 while 2014 again has a value of 5.8\%. The differences are mainly constituted by the different demands for the energy carriers that are used to supply the space heat, while the electricity demands remain almost constant for all periods. It seems that in relatively mild weather years, the deviation is higher than in colder weather years. A probable explanation of the varying deviations could be an adaptive occupancy behavior, e.g., the ventilation rates could be reduced in colder winters, which is not taken in to account in the model. 

Figure~\ref{fig:EnergyDemandYearVarSpatial} shows the spatial distribution of the final energy demand averaged for all of the weather years considered. It clusters in the cities as expected. Additionally, the relative changes of the final energy demand for the different weather years are illustrated for the different municipalities. The overall magnitudes of the differences align with the differences shown in Figure~\ref{fig:EnergyDemandYearVar}. Nevertheless, it is clearly recognizable that different weather years spatially impact the annual energy demand spatially differently: While the year 2010 was generally a cold year, in northern Germany the final energy demand was 17\% above the average while in south-west Germany it was only 11\% higher than the average. This is the opposite to 2013, when Southern Germany's energy demand was 9\% above the regional average, while northern Germany lies just 5\% above the average. This highlights the importance of having a spatially resolved building stock model, as a single location is not able to represent this variation sufficiently.


\myHereWidthFigure[fig:EnergyDemandYearVarSpatial]{images//EnergyDemandYearVarSpatial}{Spatial distribution of the final energy demand, averaged for the considered years 2010 until 2015, and the relative regional deviation from the average value.}{Final energy demand for different years}%

In 2014, no significant differences due to the geo-position are observed. Nevertheless, it becomes clear that the cities are less sensitive to the weather patterns (11.5\% below the average in 2014) than the rural areas (13.5\% below the average in 2014). The reason is that the relative share of energy demand for space heating to the overall energy demand is smaller in the cities than in the rural areas, reducing the relative impact of weather years on the total energy demand. 

The analysis illustrates that the novel spatially resolved approach is able to identify local extreme weather patterns. While it was only shown here for the aggregated annual demand, the model also predicts the temporal demand of the energy carriers in all municipalities and can be used for the identification of local peak demands that are relevant for the infrastructure's design.

%% file: 04_Results.tex
\section{Future energy supply scenario}
\label{sec:res_mincost}

In order to predict the change of the supply structure in the future, the overall model is applied with 200 archetype buildings for the techno-economic assumptions in 2050, which are defined in \ref{sec:res_assump_2050}. The choice, scale, and operation of the considered energy supply technologies are optimized together with the heating system and potential refurbishment measures, implying that the building owners have a technology adoption and operation approach that minimizes their energy cost and act as \textit{homo economicus}. We define this as \textit{Min Cost} scenario.

The results define the overall state that the residential energy supply system is converging on if the assumed energy prices and the techno-economic assumptions for the technologies arise. Besides the incentives included in the scenario, no additional ones are given by regulators. The demand for the use of electrical devices, hot water demand and thermal comfort level are asumed to stay the same as the status quo described in Section~\ref{sec:Validation} with the assumptions described in~\ref{sec:res_assump_2050}. Existing technologies are assumed to get replaced until 2050, wherefore each building has a greenfield optimization regarding the supply side.


\subsection{Design, costs and operation}
\label{sec:res_mincost_cost}

The resulting total expenditures related to investments in the supply structure or energetic refurbishment measures are visualized in Figure \ref{fig:MinCostInvestment0}. The aggregated capacities and energy flows are shown in \ref{sec:results_tables}. To realize the technology portfolio, an overall investment of 382.3 billion Euro is needed. The largest share is photovoltaics, with a total investment of 104.6 billion Euro and a total capacity of 133.4 GW. 89.7 TWh/a of the generated electricity is used for self-consumption, which is the main incentive to deploy photovoltaic. 

The second highest investment is for heat pumps with 88 billion Euro indicating that these are the main supplier of space heating. Fuel cells are the chosen flexible co-generation option and amount to 5.4\% of the annual costs. The heat storage systems make up 1.9\% of the annual costs and have a total investment of 21.8 billion Euro. The investment in the batteries is significantly lower with 6.9 billion Euro, amounting to 1\% of the annual costs. The log wood supply for the fireplaces amounts to 2.1\%, while the electric heaters have a minor share. District heating, oil boilers and pellet boilers are not chosen in the solution, as they are not competitive in comparison to the heat pumps or gas boilers. 


\myHereWidthFigure[fig:MinCostInvestment0]{images//MinCostInvestment0}{Total investments into the different measures in the residential buildings for the \textit{Min Cost} scenario in 2050. }{Total investments of the \textit{Min Cost} Scenario for Germany}%

Together, the energy-related refurbishment measures account for 97.2 billion Euro while more than half of them are determined by the walls. The occupancy control systems have a relatively high share with 12.8 billion Euro, followed by the windows and the roofs. One reason for the relatively small cost share of the efficiency measures is that most measures are chosen for buildings that are anyway in the refurbishment cycle due to their lifetime. Therefore, the costs for the sole efficiency measures are relatively small because installation costs, such as scaffolding construction, are seperately considered for these buildings. 

The overall results are aggregated from the optimal system design of the different archetype buildings, whose cost structure is illustrated in Figure \ref{fig:minCost_TRY_default_2050_0}. The total annual cost of the buildings is scaled by the number of households in the buildings to show different sizes of buildings on a similar scale. In order to expose patterns between the buildings, they are manually clustered into four groups based on their resulting supply system. The Single Family Houses (SFHs) are manually differentiated between those with and those without heat pumps, while the Multi Family Houses (MFHs) are distinguished between those with and without fuel cells.

\myHereWidthFigure[fig:minCost_TRY_default_2050_0]{images//minCost_TRY_default_2050_0}{Cost composition of the different archetype buildings for the \textit{Min Cost} scenario in 2050. They are grouped by Single-Family House (SFH) with and without (wo.) heatpumps and Multi-Family Houses (MFH) with and without (wo.) fuel cells.}{Cost composition of the archetype buildings}%

In general, the only technology that is chosen for almost all of the buildings is rooftop photovoltaic. With the predicted small cost of the photovoltaic panels and high electricity price, these are in the cost-optimal solution for various scales but independent of the roof orientation of the building. 

Except for one SFH that has a completely self-sufficient electricity supply, no other SFH has a fuel cell installed. Since the demand profile of a single-family house is highly volatile, the achievable full load hours for a self-sufficient electricity supply are too low for a fuel cell to become economically feasible. Furthermore, the required capacities of the fuel cell would be small, increasing the specific cost due to a missing economy of scale.

Moreover, it is striking that the occupancy controllers are primarily installed in SFHs with gas boilers. The building cluster with gas boilers is dominated by compact buildings where only a few rooms need to be equipped with the thermostats, constituting small investment costs. Moreover, the heat capacity of those buildings is small and constitutes limited thermal inertia. This is beneficial for the occupancy controller, as the building can cool down and heat up faster in the case of vacant occupants. For large buildings with a high thermal mass, an occupancy controller only offers limited benefit.

All MFHs with a fuel cell have an additional heat pump installed. The cheap self-supply with electricity benefits electrical heat generation. Some of the MFHs add a battery system to increase the share of the photovoltaic and CHP electricity that can then be self-consumed.


The different full load hours and capacities of the technologies in the different archetype buildings are shown in Figure \ref{fig:results_mincost_techload}. The scale of the dots indicate how often the archetype buildings are assigned in total in Germany. In general, it can be seen that although photovoltaics are installed in all buildings, the achievable full load hours vary from 683 to 1025 depending on the roof orientation and location of the archetype building. 

The highest full load hours are around 5000, and achieved by the fuel cells. It is observed that a larger fuel cell capacity correlates with higher achievable full load hours. This mainly relates to the occupancy profiles: due to statistical balancing effects, larger buildings have flatter profiles that can be covered by higher self-generation rates. Opposing effects are observed for the peak generators, such as the gas boiler with around 2000 full load hours and the electric heater with less than 1000 full load hours: the larger the installed capacities are, the smaller are the achievable full load hours. For the heat pump, no such effect is observed. It is operated with between 3000 and 4000 full load hours for small capacities as well as large ones.

\myHereWidthFigure[fig:results_mincost_techload]{images//minCost_0_TechLoad}{Full load hours and capacity of the installed technologies in the different archetype buildings for the \textit{Min Cost} scenario in 2050. The size of the scatter is related to the overall appearance of the archetypes in Germany.}{Full load hours and capacities of the technologies.}%

The distribution of scales and full load hours indicates that the heat pumps significantly rely on a peak boiler, since their scaling to the maximal heat load would be more expensive. Nevertheless, it is open as to which peak boiler is chosen in the model. For a few full load hours, the electric heater is more cost-effective, while for many peak load hours an investment into a gas boiler could be advantageous. From a central infrastructure perspective, both options have an intrinsic economic issue, as they need the layout of an infrastructure that will be used in its maximal capacity for only a few hours per year.

\subsection{Changing electricity grid load}
\label{sec:res_mincost_load}

The resulting electricity grid exchange, defined by the electricity imported for the heat pump, the conventional electricity demand, and the photovoltaic feed-in is illustrated in Figure \ref{fig:MinCostElectricityLoad}. For comparison purposes, the grid exchange of the \textit{Reference} scenario is shown as well. The aggregated electricity load of the \textit{Reference} scenario is dominated by the occupant activities in the morning and evening. A small variation between winter and summer then appears. The overall load peaks in the evening hours during winter with 36.4~GW. This aggregated load significantly changes for the 2050 scenario, when during the summer the load demand is reduced to values below 10~GW, and also for the evening hours, while high daytime feed-in rates of the photovoltaic occur with up to 43.1~GW, exceeding the peak demand of the \textit{Reference} scenario. The impact of the photovoltaic gets reduced during the winter but still reduces the load at noon for most days. The evening hours in winter are still the peak demand times with a load of up to 32.3~GW for the 2050 scenario, which is in a similar in magnitude to the \textit{Reference} scenario.

\myHereWidthFigure[fig:MinCostElectricityLoad]{images//MinCostElectricityLoad}{Aggregated grid exchange of the national residential building stock for the \textit{Reference} and the \textit{Min Cost} scenario.}{Grid exchange of the \textit{Reference} and \textit{Min Cost} scenario in 2050}%

As introduced previously, the technology installations depend on the building type. Therefore, the changes in the grid load also vary spatially depending on the local building topology, as illustrated in Figure \ref{fig:MinCostLoadChangeSpatial}. As expected, the majority of the regions reduce their annual electricity demand with the help of self-generation by photovoltaics and fuel cells. Nevertheless, regional differences are high: while urban areas are able to reduce their electricity demand by 60\%, some rural areas even increase it. The high photovoltaic installations in rural areas are not sufficient to compensate for the increased electricity demand from the heat pumps. This effect intensifies for the case of the peak load, as almost no photovoltaic feed-in exists in the winter days, while the heat pumps are being operated in full load. Therefore, regions characterized by large SFHs double their peak load. This is different for the urban areas that even reduce their peak load because the fuel cells exceed the electrical capacity of the heat pumps and are synchronously operated. Equivalent regional trends are observed for the feed-in: The rural areas feed up to 40\% of the original electricity demand into the grid, while the urban areas have only small feed-in rates of 10\%. Further, a gradient between north and south is recognizable due to the different solar irradiance.

\myHereWidthFigure[fig:MinCostLoadChangeSpatial]{images//MinCostLoadChangeSpatial}{Spatial change of the peak electricity demand and the change of the cumulative positive demand from the \textit{Reference} scenario to the \textit{Min Cost} scenario. Furthermore, the amount of electricity feed-in to the grid in the \textit{Min Cost} scenario is shown in ratio to the cumulative electricity demand in the \textit{Reference} scenario. }{Spatial change of grid load from \textit{Reference} to \textit{Min Cost} scenario.}%



In summary, the results indicate that the change of the energy supply in the rural areas is more challenging with respect to the electricity grid operation than the changes in the urban areas. Nevertheless, adaptions in the tariff design could dampen this effect. 

\subsection{Value of analysis}
\label{sec:res_mincost_valueof}

In order to evaluate the robustness of the \textit{Min Cost} scenario for 2050,  200 archetype buildings are again optimized, but parts of the technologies are excluded or forced into the solution space.

Figure \ref{fig:ValueOfAnnualCost} illustrates the aggregated resulting annual cost composition of the different cases that were considered for the analysis. Gas supply, fuel cell, photovoltaic, heat pump and refurbishment measures are each excluded from the solution space, and the other cost minimal solutions are compared to the original \textit{Min Cost} scenario with all technologies available. The increase in the total systems costs can be interpreted as the \textit{Value Of} the integration of a certain technology. As an additional case, the full package of refurbishment measures are enforced for all buildings that are in the refurbishment cycle in order to reach lower demands for space heating.

\myHereWidthFigure[fig:ValueOfAnnualCost]{images//ValueOfAnnualCost}{Annual cost of the \textit{Min Cost} scenario and the resulting aggregated system cost if the solution space is constrained.}{Cost composition of the Value of analysis}%

In case the fossil gas supply is excluded from the solution space, the electricity purchase doubles, as no fuel cells for self-consumption are installed. Bio-methane or another renewable fuel is too expensive in the considered scenario to replace the fossil gas in the fuel cells. Instead, higher capacities of photovoltaics are integrated into the solution, aggregating up to 160.3~GW. Moreover, the aggregated cost for heat pumps increases by 38\% as their share of the heat supply increases. While the \textit{Min Cost} solution did not include district heating or pellet boilers, they are used in small scales in the case that fossil gas is excluded. The amount of occupancy controllers is also reduced without fossil gas. The reason is that the heat pump is intensively used during the day in order to utilize photovoltaic electricity while heating up the building. Nevertheless, the occupancy controller lowers the comfort temperature especially during the day when the occupants are absent. These two temporally opposing effects reduce the value of an occupancy controller for the buildings supplied with heat pumps. The exclusion of fossil gas also constitutes the smallest GHG footprint that gets reduced from 51.9 Mt/a in the \textit{Min Cost} scenario to 19.9 Mt/a, as shown in Figure \ref{fig:ValueOfCO2emissions}.

\myHereWidthFigure[fig:ValueOfCO2emissions]{images//ValueOfCO2emissions}{GHG footprint of the \textit{Min Cost} scenario and the resulting aggregated GHG footprints if the solution space is constrained.}{GHG footprints of the Value of analysis}%

The structural changes to the \textit{Min Cost} scenario are fairly small for the case that the fuel cell is excluded from the solution space. The net electricity import increases as in the previous scenario and compensates for the missing self-generation, but the photovoltaic capacities merely increase from 133.4~GW to 142.3~GW, while the heat pump capacities remain in a similar magnitude. This is different to the previous case and indicates that the value of further photovoltaic capacities is mainly correlated to higher heat pump capacities and not to smaller fuel cell capacities. The battery capacities are reduced from 16.9~GWh to 12.8~GWh, although the photovoltaic capacity is increasing. This implies that their operation partially complements the fuel cell operation. 

Significant shifts and cost increases are recognizable in the case that the photovoltaic is excluded, while the electricity purchase only increases from 71.9~TWh/a to 80.9~TWh/a, the gas import almost doubles from 172.5~TWh/a to 285.7~TWh/a with the effect of a GHG footprint of up to 81.3 Mt/a. High gas boiler capacities compensate for the reduction of the heat pump capacities from 60.4 to 46.9~GW$_{th}$. This indicates the enforcing effect between the heat pump and the photovoltaic supply, which is economically advantageous in the case of self-consumption with photovoltaics is available. No battery capacities are installed, supporting the statement that their main economic driver for installation is the photovoltaic, although they are also partially used to increase the self-consumption with fuel cell electricity.

In the case that a heat pump is excluded from the solution space, the amount of gas increases by 113.2~TWh/a while the electricity demand only gets reduced by 13.5~TWh/a. In consequence, the GHG footprints increase by up to 76.0 Mt/a. Furthermore, the investment in refurbishment measures increases by 30\%, dominated by more occupancy controller and more wall insulation and vice versa, indicating that especially cheap heat produced by the heat pumps lowers the motivation to invest in efficiency measures. Also, the fire wood supply increases from 21.2~TWh/a to 51.44~TWh/a, since it is a cheaper fuel than fossil gas in the scenario. Remarkable is the reduced investment in fuel cells, cutting their capacity from 12.7 to 5.1~GW$_{el}$. This illustrates that major fuel cell capacities are built to supply the heat pumps with electricity.

The exclusion of refurbishment measures from the solution space constitutes an increased investment in heat pumps and a reduced investment in gas boilers. This is surprising, as an enforcing effect between the heat pump and refurbishment measures could be expected because the refurbishment measures decrease the required supply temperature in the building, and vice versa, increasing the efficiency of heat pumps. Nevertheless, the economic effects predominate, the heat pumps have higher investment costs than the gas boilers, while on the other hand the operational energy cost for the gas supply is higher. In consequence, heat pumps are favored in the case of high heat demands and their deployment increases for the case of no refurbishment. 

\myHereWidthFigure[fig:ValueOfCostIncrease]{images//ValueOfCostIncrease}{Annual cost increase for the case that certain technologies are excluded or added to the solution space in the \textit{Min Cost} for 2050. The change is once shown for the total aggregated cost and once for the single archetype buildings.}{Distribution of the cost increase in the Value of analysis}%

The reverse effect occurs for the forced refurbishment case: Installed heat pump capacities are reduced while gas boiler capacities increase. The overall demand for gas and electricity is reduced as the space heat demand drops to 209.1~TWh/a, in comparison to the 309.8 TWh/a in the \textit{Min Cost} scenario and the 449 TWh/a in the \textit{Reference} scenario. Nevertheless, the demand reduction is not able to compensate for the high cost of the refurbishment measures, resulting in an overall annual cost increase of 29.7\%. In particular, ventilation systems with heat recovery amount to almost half of the efficiency measure costs. Ventilation systems do not benefit from the refurbishment cycle, as their integration cost in the building is mostly independent of any outside renovation measures. Also noticeable is the fact that the amount of occupancy control systems drops: If the heat demand is reduced anyway, additional measures such as temporally reducing the inner air temperature have a minor effect, making the occupancy controller economically unfavorable. The GHG footprint is only reduced to 45.1 Mt/a, which is much smaller than expected but explained by the switch of many buildings to gas boilers. This indicates a potential rebound effect that might occur in the future, in that the reduced demand for space heating lowers the economic incentive to invest in efficient but expensive heat supply technologies. 

It is striking that the aggregated cost gain is moderate for all considered cases, except for the forced refurbishment case. This indicates that the prediction of the total cost is robust and not sensitive to the available technologies in the future, although, this robustness only accounts for an aggregated German-wide level, as illustrated in Figure \ref{fig:ValueOfCostIncrease}. The figure shows the total cost increase  and the distribution of the cost increase of the single buildings. While the total cost in Germany only increases by 3.65\% for the case that no fossil gas supply is available, one of the archetype buildings has a 22.9\% higher energy costs, while some other buildings are not affected at all, as they were also not supplied with gas in the \textit{Min Cost} scenario. Similar effects are observed for the other sensitivity analyses: The sensitivities for single buildings are high, but the cost of the aggregated result is robust. 


The impact on the electricity grid of the different cases is further illustrated in Figure \ref{fig:ValueOfSortedElec}. It shows the sorted grid load for the \textit{Reference} scenario, the \textit{Min Cost} scenario, and all related sensitivity analyses. The highest peak load occurs if the gas supply is completely excluded from the solution. No significant gas boiler capacities are able to satisfy the peak heat demand and no fuel cells can diminish the additional electricity load of the heat pumps. In consequence, the peak load almost doubles to 55.9~GW in comparison to the \textit{Min Cost} scenario with 32.3~GW. The second highest demand is reached if no fuel cell is included and the peak load increases by 14.6~GW relative to the \textit{Min Cost} scenario. This shows the importance of decentralized flexible electricity generation in order to compensate for the increased electricity demand by the heat pumps. As is to be expected, the case without heat pumps has the lowest peak load, with 25.7~GW. 

\myHereWidthFigure[fig:ValueOfSortedElec]{images//ValueOfSortedElec}{Sorted grid load of the \textit{Min Cost} scenario and the grid load if the solution space is constrained.}{Sorted grid load Value of analysis}%

The amount of photovoltaic feed-in is for all cases that include photovoltaics in a similar range. The maximal feed-in is reached with 55.4~GW for the exclusion of the gas supply, although the single buildings have the constraint to limit the feed-in to 50\% of their maximal capacity.


%% file: 05_Conclusion.tex
\section{Limitations and outlook}

As the previous sections showed the capability of the modeling approach, it has also some limitations.

\subsection{Chosen archetypes}

As concluded by \citet{Corgnati2013}, the aggregation and descriptions of the archetype buildings is highly dependent on the available data. The algorithm in this work is tailored for the data structure of the German Census \citep{Statistisches2011} that describes the statistical distributions of building parameters at the municipality level. The methodology is transferable to other countries where similar data structure exists, but in case that exact building samples are available, it would still be recommended to use conventional cluster algorithms for the aggregation. The fitting of archetype buildings to meet attribute distributions, as in this work, has the drawback that theoretical building configurations are created that meet the distribution values but can significantly deviate from real building instances. 

Further, while the set of archetype buildings is able to respect the variety of households with different cumulative electricity demands due to their different appliance equipment and different household sizes, the appliance equipment and adoption rate are not altered by the socioeconomic background of the households. Nevertheless, this is a significant descriptor to determine electricity demand variation \citep{Druckman2008,Elnakat2016} and probably also technology adoption rates. The approach would allow this description, but the required data is not publicly available. 

For more holistic analysis, the algorithm should be transferred to other sectors to derive spatially-distributed sector specific representatives. Thus, a cross-sectoral spatially resolved bottom-up model can be derived that respects the individual economic entities. Examples are the service sector including commercial buildings, or also representative fueling stations, whose detailed models could be upscaled to a nationwide perspective while respecting the spatially varying conditions to supply them. 


\subsection{Buildings as economic entities}

A limitation is the current scope of the building model: Mobility demand should be included in future since it has also a high impact on the grid load and provides further flexibility. Thereby, also further technology should be included, such as ground-source heat pumps or hydrogen storage technologies. While those technologies can be easily added, the single archetype approach does not allow a good evaluation of district scale technologies since the entities are considered independently and occure in the municipalities in different combinations, wherefore a systematic evaluation, e.g. of heating networks, is hardly possible.

Supplementary, the sole financial agent decision making is known to perform well in rate of adoption and cumulative adoption but underestimates social and attitudinal components influencing the technology adoption \citep{Robinson2015}. In consequence, only a few economically dominant technologies are chosen in the \textit{Min Cost} case, although even a higher diversity would exist in reality due to individually varying information levels. As a case in point, pellet boilers were down-selected, although those are considered in different scenarios in the literature \citep{IWES2015,BMWi2015}. This dominance is related to the scenario, and a consideration of different biomass prices could change this. Nevertheless, in reality an adoption of pellets would also be expected without being economically competitive. Such non-cost optimal adoption behaviors are better included in simple adoption models such as \textit{Invert} tool \citep{Kranzl2013,Muller2015} that includes a statistical randomness in the adoption process, but neglects, e.g., the temporal operation. Nevertheless, the information basis for investment decisions is improving wherefore the assumption in this work of a cost optimal investment stays reasonable.

\subsection{Further scenarios}

The good alignment with the reported final energy demand values of today makes the model further suitable to develop transformation paths of the building stock until the year 2050. E.g., additional projected supply years in 2020, 2030 and 2040 could be modeled. Such an approach could better consider the different lifetimes of the technologies and respect the inertia in the adoption process of the building owners.

Thereby, also short term policy design could be better supported. The resulting grid load for the 2050 scenario in this work is highly related to the flat energy tariffs considered. Nevertheless, the peak and feed-in values can be controlled by the different tariff structure, that could give incentives to shift consumption in time in order to flatten the load profiles. Therefore, upcoming works will apply the model to different regulatory regimes and market environments. A coupling with grid models is promising in order to determine an economically optimal market design.

\section{Conclusions}
\label{sec:Conclusion}

In this paper, a novel bottom-up model was introduced that is based on an aggregation of archetype buildings and a related optimization model to predict the a spatially-resolved technology adoption and operation. 

The model approach allows the evaluation of the influence of regulatory decisions on energy cost, green house gas emissions, or grid load under the assumption of cost optimal behavior. The novelty is that it is able to analyze the impact regulatory regimes and market environments for single  buildings, simultaneously with a nationwide economic perspective.

As show case, the model was applied and validated for the residential building stock of Germany and a future scenario frame for 2050. The main techno-economic conclusions drawn from the future scenario are the following:
\begin{itemize}
    \item The key technologies for reducing the GHG emissions in the building stock are photovoltaic and heatpumps that will significantly increase the seasonal variation of the residential electricity load, as their feed-in and demand do not temporally match, resulting in a doubling of the peak electricity load in the winter hours in rural areas.
    \item The urban areas can compensate the increasing electricity demand by efficient co-generation units, e.g., in form of fuel cells, which cannot achieve sufficient economies of scale in single family houses in the rural areas.
    \item Significant amounts of photovoltaic electricity (for Germany up to 90 TWh/a) can be self-consumed while the majority is used for internal heat supply applications. Batteries are hardly deployed, as the heat applications provide enough flexibility for self-consumption.
    \item Refurbishment measures are expensive and only chosen in cases when the building is in the cosmetic refurbishment cycle. Therefore, space heat demand only decreases by 30\% from 2015 to 2050 in the scenario. Instead, a shift towards an efficient integrated energy supply system, e.g. combinations of fuel cells with heat pumps, is favored.
\end{itemize}

%% file: 00_Appendix.tex
\section{Techno-economic assumptions}
\label{sec:res_assump}

The main input assumptions to parametrize the optimization models are introduced in the following section, e.g. residential energy prices and efficiencies of the different technologies. Section~\ref{sec:res_assump_2015} defines the parameters for the year 2015, while ~\ref{sec:res_assump_2050} extends and adapts them to the year 2050.

\subsection{Assumptions for 2015}
\label{sec:res_assump_2015}

In order to achieve a valid comparison of today's residential energy supply to the changes that will occur in the future, a valid scenario framework is introduced that represents today's cost and operation parameters for the residential energy supply systems. 


The economic parameters considered for the supply technologies are illustrated in Table~\ref{tab:eco_parameters_validation}, while their detailed derivation is discussed in \citet{Kotzur2018}. The structure of the investment costs is oriented around the cost model introduced in \citet{Lindberg2016}. It differentiates between the fixed investment costs that occur in the case of installation and the specific investment costs that are added and related to the scale of the installations. 


\myScaledTable{
\begin{tabular}{lrrrrp{2.5cm}}
\toprule
Technology & CAPEX & CAPEX & OPEX & Lifetime &                     Source \\
{} &         fix &   specific & \%CAPEX/a &        a &                                   \\
\midrule
Gas boiler      &        2800 Euro &        100 Euro/kW$_{th}$ &      1.5 &       20 &      \citep{Kotzur2018}  \\
Oil boiler      &        2800 Euro &        100 Euro/kW$_{th}$ &      1.5 &       20 &      \citep{Kotzur2018}  \\
Pellet boiler   &       10000 Euro &        300 Euro/kW$_{th}$ &      3.0 &       20 &    \citep{Kotzur2018}  \\
Heat pump       &        5000 Euro &        600 Euro/kW$_{th}$ &      2.0 &       20 &        \citep{Kotzur2018}  \\
Heat storage    &         800 Euro &       1200 Euro/m$^3$     &      0.0 &       25 &     \citep{Kotzur2018}  \\
Photovoltaic    &        1000 Euro &       1400 Euro/kW$_{el}$ &      1.0 &       20 &     \citep{Kotzur2018} \\
IC CHP          &       15000 Euro &       1000 Euro/kW$_{el}$ &      7.0 &       15 &          \citep{Kotzur2018}  \\
Solar thermal   &        4000 Euro &        350 Euro/m$^{2}$   &      1.0 &       20 &    \citep{Kotzur2018}  \\
Electric heater &           0 Euro &         60 Euro/kW$_{th}$ &      2.0 &       30 &  \citep{Lindberg2016} \\
\bottomrule
\end{tabular}
}{Assumed economic parameters of the energy supply technologies for the \textit{Reference} scenario.}{tab:eco_parameters_validation}{Economic parameters of the energy supply technologies for the \textit{Reference} scenario}%



Although the model allows for the modeling of different interest rates for different building types to take into account of the different investment behavior of the building owners \citep{Schleich2016}, it is here simplified to a single interest rate of 3\%, which lays between the 2\%~to~5\% considered in the literature \citep{Klingler2017,Lahnaoui2017,Lauinger2016,Lindberg2016,Schutz2017}. 


The energy and resource prices are illustrated in Table~\ref{tab:eco_supply_validation}. The majority of the prices are derived from the study \textit{Energieeffizienzstrategie Gebäude} \citep{BMWi2015,Prognos2015}. Their assumptions define the basic scenario framework for this thesis and rely themselves on the \textit{Energiereferenzprognose} \citep{EWI2014}. The majority of the resource prices assumed in the study align with the energy prices observed for 2016 \citep{Bundesnetzagentur2017,BMWi2018}. Nevertheless, the assumed gas price overshoots the observed price of 2016 by more than 1 ct/kWh, and was adapted in this work to the values reported for 2016 by the \citet{Bundesnetzagentur2017}. 

The GHG footprint includes the emissions of the previous conversion processes, such as in the extraction of fuels, or the GHG emissions of the German power plant mix. 

\myScaledTable{
\begin{tabular}{p{2cm}rrrrp{3cm}}
\toprule
Technology & OPEX-var & OPEX-fix & GHG & PE &                         Comment \\
- &  Euro/kWh &    Euro/a &        kg/kWh &              kWh/kWh &                               - \\
\midrule
Electricity supply &    0.246 &      170 &         0.525 &            1.8 &         0.292 Euro/kWh for 3700 kWh/a \\
Gas supply         &    0.065 &        0 &         0.250 &            1.1 &                         \\
Oil supply         &    0.064 &        0 &         0.320 &            1.1 &                         \\
Pellet supply      &    0.060 &        0 &         0.014 &            0.2 &                                \\
Heat pump tariff           &    0.190 &       70 &         0.525 &            1.8 &                                \\
FiT CHP             &   -0.08 &        0 &         0.000 &            2.8 &               for  less than 50 kWel \\
FiT PV              &   -0.108 &        0 &         0.000 &            1.8 &                                \\
District heating   &    0.074 &      327 &         0.270 &            0.7 &  			0.096 Euro/kWh for 15.000 kWh/a \\
Log supply         &    0.050 &        0 &         0.000 &            0.2 &                                \\
\bottomrule
\end{tabular}
}{Assumed residential energy prices including taxes, levies, and network charges based on \textit{Energieeffizienzstrategie Gebäude} \citep{EWI2014,Prognos2015,BMWi2015} while missing parameters are derived from \citet{Lindberg2016a,KWKG2016,EEG2017}. The gas price is corrected to the observed gas prices in 2016 \citep{Bundesnetzagentur2017}. The GHG footprint and primary energy factors (PE) are taken from \citet{Prognos2015}. FiT refers to Feed-in Tariff.}{tab:eco_supply_validation}{Assumed residential energy prices and GHG emissions for the \textit{Reference} scenario}%

Furthermore, the price structure is modified from a sole energy price (Euro/kWh) structure to a combination of a flat price (Euro/a) and an energy price (Euro/kWh). This is important because the savings due to self-consumption, e.g. of photovoltaic electricity, would be overestimated with the sole energy price. Additionally, this structure respects that specific wholesale prices decrease with larger energy consumptions rates \citep{Bundesnetzagentur2017}. 

The technical performance of the technologies is summarized in Table~\ref{tab:tech_parameters_validation}. The efficiencies are given for the Lower Heating Value (LHV) of gas, oil or pellets. The electrical and thermal CHP efficiencies are defined for a fixed operation ratio and cannot be varied in between. The values are chosen such that the different age structures of the technologies are respected, e.g. an efficiency is assumed for the gas boiler that refers to the efficiency of condensing boilers, while for the oil boiler a lower efficiency is considered that is related to older boiler technologies.

\myScaledTable{
\begin{tabular}{lll}
\toprule
Technology & Efficiency &      Comment and \textit{Reference} \\
\midrule
Gas boiler      &        0.96 		&      Condensing boiler   \\
								&       				 	&        \citep{Henning2014} \\
Oil boiler      &        0.84  		&  \citep{UBA2017a}  \\
Pellet boiler   &       0.9 			&       \citep{Lindberg2016} \\
Heat pump       &        dynamic  &   \citep{Kotzur2018} \\
								&        				  &     quality grade of 0.4 \\
Heat storage    &         0.99 		&      charge \citep{Lindberg2016} \\
								&         0.99 		&       discharge \\
								&         0.6\%/h &  self-discharge	\citep{Schutz2015} \\
Photovoltaic    &        0.15 	& based on Hanwha HSL 60 S \citep{King2016} \\
								&         				&  with 7 m$^2$/kWp \\
IC CHP         &       0.6 			&   thermal  \citep{ASUE2015}  \\
								&       0.25 			&   electric  \citep{ASUE2015}  \\
Electric heater &       0.98 			& \citep{UBA2017a} \\
Solar thermal   &        dynamic 	&  \citep{Lindberg2016}  \\
Fireplace     &        0.83 & \citep{UBA2017a,Olsberg2018} \\
\bottomrule
\end{tabular}
}{Summary of the main technical parameters of the energy supply technologies}{tab:tech_parameters_validation}{Technical key parameters of the energy supply technologies}%

The comfort temperature inside the buildings is assumed to have a value of 21°C for when occupants are active at home. The night reduction temperature is set for all buildings to 18°C.

\subsection{Assumptions for 2050}
\label{sec:res_assump_2050}

The techno-economic assumptions for the future energy supply through 2050 are introduced in the following section. While many parameters are estimated to stay at a similar magnitude as in the \textit{Reference} case in Section~\ref{sec:res_assump_2015}, this section describes only the assumptions that are changing for the case of 2050. All prices and costs are provided as real prices in 2015.

While no major changes are expected for conventional heat generators, further learning rates and cost reductions are considered for photovoltaic and electrochemical technologies, as shown in Table~\ref{tab:eco_parameters_2050}. Their detailed derivation and discussion is also performed in \citet{Kotzur2018}. 

\myScaledTable{
\begin{tabular}{lrrrrp{2.4cm}}
\toprule
Technology & CAPEX & CAPEX & OPEX & Lifetime &                     Source \\
{} &         fix &   specific & \%CAPEX/a &        a &                                   \\
\midrule
Photovoltaic &         1000 Euro &         650 Euro/kW$_{el}$ &      1.0 &       20 &     \citep{Kotzur2018} \\
Battery			 &         1000 Euro &         300 Euro/kWh  			&      2.0 &       15 &    \citep{Kotzur2018} \\
Fuel cell		 &         4000 Euro &        1500 Euro/kW$_{el}$ &      3.0 &       15 &     \citep{Kotzur2018} \\
\bottomrule
\end{tabular}
}{Change and addition of economic parameters of the energy supply technologies for the year 2050.}{tab:eco_parameters_2050}{Economic parameters for the supply technologies in the year 2050}%

The technical assumptions for 2050 are shown in Table~\ref{tab:tech_parameters_2050}. The efficiency of the heat pumps is expected to increase further in the future \citep{Willem2017}, for which this work assumes an increase of the quality grade to 0.45, which is the upper bound of today's systems \citep{Kotzur2018}. The photovoltaic efficiency is assumed to increase to a value of 30\% \citep{ISE2015}. Primarily, this impacts the space coverage on the rooftop and increases the photovoltaic potential that can be installed. The technical parameters of the batteries are derived from a prediction until 2050 \citep{Elsner2015}, but some of today's residential storage systems already achieve similar efficiencies \citep{Kairies2016}.

\myTable{
\begin{tabular}{lll}
\toprule
Technology & Efficiency &      Comment and \textit{Reference} \\
\midrule
Heat pump       &        dynamic &     \citep{Kotzur2018} \\
								&        				 &      quality grade of 0.45 \\
Photovoltaic    &        0.3 & average 2050 \citep{ISE2015} \\
								&         &  with 3.5 m$^2$/kWp \\
Battery			    &         0.95 		&      charge \citep{Elsner2015} \\
								&         0.95 &       discharge \citep{Elsner2015} \\
								&         0.01\%/h &  self-discharge	\citep{Elsner2015} \\
								&         0.5 kW/kWh &  capacity factor \\
Fuel cell         &       0.33 &   thermal  \citep{Kotzur2018} \\
								&       0.52 &   electric  	\citep{Kotzur2018}   \\
\bottomrule
\end{tabular}
}{Summary of the main technical parameters of the energy supply technologies for 2050. }{tab:tech_parameters_2050}{Key technical parameters of the supply technologies in 2050}%

The electrical efficiency of the fuel cell is assumed to be 52\% and positions itself between the efficiency that can be achieved from Solid Oxide Fuel Cell (SOFC) systems and the efficiency of the Proton Exchange Membrane Fuel Cells (PEMFC), as discussed in detail in \citet{Kotzur2018}. A fully flexible operation is assumed for the year 2050. The efficiencies are considered to be the same for natural gas, biogas or hydrogen as potential alternative fuels \citep{Peters2016}.

The energy prices for 2050 are shown in Table~\ref{tab:eco_supply_future} and also rely on the \textit{Energieeffizienzstrategie Gebäude} \citep{BMWi2015,Prognos2015} and \textit{Energiereferenzprognose} \citep{EWI2014}. The \textit{Energiereferenzprognose} considers a carbon price of 76 Euro/ton for the year 2050, which, e.g., increases the gas price by 1.9 ct/kWh.

\myScaledTable{
\begin{tabular}{p{2cm}rrrrp{3cm}}
\toprule
Technology & OPEX-var & OPEX-fix & GHG & PE &                         Comment \\
- &  Euro/kWh &    Euro/a &        kg/kWh &              kWh/kWh &                               - \\
\midrule
Electricity supply &    0.220 &      170 &         0.122 &            0.4 &         0.266 Euro/kWh for 3700 kWh/a \\
Gas supply         &    0.096 &        0 &         0.250 &            1.1 &           				 \\
Bio-methane      &    0.138 &        0 &         0.014 &            0.2 &                   \\
Oil supply         &    0.124 &        0 &         0.320 &            1.1 &           				 \\
Pellet supply      &    0.080 &        0 &         0.014 &            0.2 &                   \\
HP Tarif           &    0.190 &       70 &         0.122 &            0.4 &                   \\
FiTCHP             &   -0.010 &        0 &         0.000 &            0.4 &  								 \\
FiTPV              &   -0.010 &        0 &         0.000 &            0.4 &                   \\
District heating   &    0.085 &      327 &         0.144 &            0.5 &                 0.107 Euro/kWh for 15000 kWh/a  \\
Log supply         &    0.065 &        0 &         0.000 &            0.2 &                   \\
\bottomrule
\end{tabular}}{Assumed energy prices, GHG footprints and primary energy factors (PE) based on the \textit{Energieeffizienzstrategie Gebäude} \citep{EWI2014,Prognos2015,BMWi2015} for 2050.}{tab:eco_supply_future}{Energy prices and GHG footprints in the year 2050}%

Furthermore, a bio-methane purchase is integrated with a price of 13.8 ct/kWh, which can be either a synthetic gas or biogas. As no sufficient predictions for bio-methane prices in 2050 are te be found, its price is derived from the production cost of bio-methane for the feed-in into the gas grid of 7.5 ct/kWh in 2013 \citep{Bundesnetzagentur2014}, plus the surcharge for grid fees, tax etc. This surcharge is considered to be 6.3 ct/kWh, which is the difference between the gas market price of 3.3 ct/kWh and the residential gas price of 9.6 ct/kWh in 2050 \citep{EWI2014}. All in all, it results in a price of 13.8 ct/kWh for the bio-methane, which is significantly above the fossil gas price. 


No values for future feed-in tariffs were identified. Therefore, the feed-in is only marginally subsidized, as it is highly dependent on the future market environment. A marginal value of 0.01 eur/kWh is chosen in order to guarantee that photovoltaic generation is not curtailed and is instead fed-in to the grid. 

The cost and energetic impact of the refurbishment measures for the opaque building envelope are shown in Table~\ref{tab:eco_wall}. 
\myScaledTable{
\begin{tabular}{llrrrrr}
\toprule
Component & Measure & Thickness* & Lambda & CAPEX & CAPEX energy ** \\
&- &  m &    W/m/K &    Euro/m$^2$ &  Euro/m$^2$  \\
\midrule
Wall & Base  & 	0.15 &	0.035	& 124.0 &	51.5	 \\	
		& Future & 0.22	&	0.035	& 140.9 &	68.5	\\	
\midrule
Roof & Base  & 	0.24 &	0.035	& 237.6 &	53.0	 \\	
		& Future & 0.36	&	0.035	& 270.0 &	79.6	\\	
\midrule
Floor & Base  & 	0.08 &	0.035	& 51.7 &	-	 \\	
\bottomrule
\end{tabular}}{Techno-economic assumptions for the insulation measures of a single building. The two measure levels are derived from \citet{Schutz2017} while the exact cost and lambda are taken from \citet{BMVBS2012}.(* thickness equivalent. ** capital expenditures related only to energetic measures.)}{tab:eco_wall}{Techno-economic assumptions for the insulation measures of a single building}%

All measures are additional layers to the existing envelope of the building. The costs are average values taken from a survey about subsided refurbishment measures in Germany \citep{BMVBS2012}. They differ between the entire CAPEX of a refurbishment measure and the sole additional CAPEX of energy efficiency measures if the building would have been refurbished anyway, as discussed in \citet{Kotzur2018}. The costs relate to the exterior surface area of the building component. Two levels of potential insulation measures are considered and differ by the thickness of the insulation layer, and are referred to as \textit{Base} and \textit{Future}. 

The cost for envelope refurbishment measures differs between buildings that are in the refurbishment cycle and buildings that are not, as discussed in detail in \citet{Kotzur2018}.

The costs of replacing the windows and changing the solar and thermal transmittance of the different window types are shown in Table~\ref{tab:eco_window} and rely on the \citet{BMVBS2012} as well. The costs are specific to the window area of the building. Again, a differentiation is made between the \textit{Base} and \textit{Future} levels.

\myScaledTable{
\begin{tabular}{lrrr}
\toprule
 Measure & Solar transmittance & Thermal transmittance & CAPEX  \\
 &   - &    W/m$^2$/K &  Euro/m$^2$  \\
\midrule
Base	&	0.575 &	1.1 &	313		 \\
Future	& 0.5 &	0.7	& 361.5  \\
\bottomrule
\end{tabular}}{Techno-economic assumptions for the windows. The transmittance are based on \citep{Schutz2017} and the cost based on \citep{BMVBS2012} }{tab:eco_window}{Techno-economic assumptions for the windows}%

All envelope measures have a lifetime of 40~years with zero operational costs.

In addition to the conventional refurbishment measures at the envelope of the building, a heat recovery for the ventilation is assumed with a specific investment of 65 Euro/m$^2$ per living area, a lifetime of 25 years and operational cost of 4\% per year according to the \citet{BMVBS2012} as a ratio to the original investment. If integrated, 80\% of the heat losses due to ventilation would be recovered.

Lastly, an occupancy controller can be installed that reduces the comfort temperature in case of vacant occupants \citep{Kotzur2018}. Based on the cost of \citet{Controme2018}, they are assumed with a fixed investment of 1000~Euro for the central system controller and 3~Euro/m$^2$ per living area for the different thermostats in the rooms, including their installation costs. A lifetime of 15~years is assumed. 

%
%
%
%
%
%

\section{Supplementary results}
\label{sec:results_tables}

The composition of the total annual residential energy cost for the \textit{Min Cost} scenario in 2050 scenario are shown in Figure~\ref{fig:MinCostCostShare0}.

\myHereWidthFigure[fig:MinCostCostShare0]{images//MinCostCostShare0}{Composition of the total annual costs over the whole of Germany for the \textit{Min Cost} scenario in 2050.}{Total annual costs of the \textit{Min Cost} Scenario for Germany}%

The resulting annual energy flows between the different technologies for the \textit{Min Cost} scenario in 2050 are illustrated in Figure \ref{fig:MinCostSankey0} for the aggregated level of the whole of Germany.  

\myHereWidthFigure[fig:MinCostSankey0]{images//MinCostSankey0}{Annual energy flows in~TWh between the different technologies aggregated for the whole of Germany for the \textit{Min Cost} scenario}{Energy flows in the \textit{Min Cost} scenario}%

The aggregated temporal operation of heat technologies in the \textit{Min Cost} scenario in 2050 are illustrated in Figure \ref{fig:MinCostHeatFlows0}.

\myHereWidthFigure[fig:MinCostHeatFlows0]{images//MinCostHeatFlows0}{Heat flows of the relevant heat generators to the heat node and the connected demand for space heating for Germany in the \textit{Min Cost} scenario in 2050.}{Heat energy flows in the \textit{Min Cost} scenario}%

The aggregated annual energy flows between the different considered system components are listed in Table~\ref{tab:value_of_flows} for the different scenarios in 2050.

\myScaledTable{
\begin{tabular}{lrrrrrrr}
\toprule
  &  Min  &  No &  No &  No &  No &  No &  Forced \\
&  Cost &  Gas supply &  Fuel cell &  Photovoltaic &  Heat pump &  refurbishment &  refurbishment \\
\midrule
AC Node to Battery             &       5.2 &             2.0 &            3.7 &               0.0 &            5.1 &                6.1 &                    5.2 \\
AC Node to Building            &     112.8 &           112.8 &          112.8 &             112.8 &          112.8 &              112.8 &                  112.8 \\
AC Node to Electric heater     &      40.8 &            56.1 &           43.3 &               4.6 &           48.1 &               45.0 &                   40.0 \\
AC Node to Heat pump           &      34.3 &            26.9 &           16.2 &              24.6 &            0.0 &               50.1 &                   21.0 \\
AC Node to Hot water           &       6.8 &             6.8 &            6.8 &               6.8 &            6.8 &                6.8 &                    6.8 \\
Battery to AC Node             &       4.7 &             1.8 &            3.3 &               0.0 &            4.6 &                5.5 &                    4.7 \\
CHP to AC Node                 &       0.0 &             0.0 &            0.2 &               0.0 &            0.0 &                0.0 &                    0.0 \\
Cool supply to Building        &      19.8 &            21.4 &           19.6 &              19.0 &           18.5 &               29.0 &                   31.6 \\
Electricity supply to AC Node  &      51.5 &           100.5 &           84.7 &              63.5 &           57.7 &               47.3 &                   52.7 \\
Fuel cell to AC Node           &      53.9 &             0.0 &            0.0 &              85.3 &           20.1 &               69.1 &                   45.7 \\
Gas supply to CHP              &       0.0 &             0.0 &            0.7 &               0.0 &            0.0 &                0.0 &                    0.0 \\
Gas supply to Fuel cell        &     103.7 &             0.0 &            0.0 &             164.1 &           38.6 &              132.9 &                   87.9 \\
Gas supply to Gas boiler       &      68.7 &             0.0 &           91.2 &             121.6 &          226.6 &               26.8 &                   65.7 \\
HP Tarif to Heat pump          &      20.4 &            46.2 &           40.2 &              17.4 &            0.0 &               42.1 &                   10.5 \\
Log supply to Fire place       &      21.2 &            19.4 &           23.0 &              23.6 &           51.4 &               21.3 &                   28.8 \\
Pellet supply to Pellet boiler &       0.0 &            10.3 &            0.0 &               0.0 &            3.1 &                0.0 &                    0.0 \\
Renewable gas to Gas boiler    &       0.0 &             8.7 &            0.0 &               0.0 &            0.0 &                0.0 &                    0.0 \\
Heat pump to Building          &     225.1 &           296.5 &          231.0 &             169.5 &            0.0 &              363.9 &                  126.7 \\
CHP to HNode                   &       0.0 &             0.0 &            0.4 &               0.0 &            0.0 &                0.0 &                    0.0 \\
District heating to HNode      &       0.0 &             7.9 &            0.8 &               0.0 &           15.0 &                0.0 &                    0.0 \\
Electric heater to HNode       &      40.0 &            55.0 &           42.5 &               4.5 &           47.1 &               44.1 &                   39.2 \\
Fire place to HNode            &      17.6 &            16.1 &           19.1 &              19.6 &           42.7 &               17.7 &                   23.9 \\
Fuel cell to HNode             &      34.2 &             0.0 &            0.0 &              54.2 &           12.7 &               43.9 &                   29.0 \\
Gas boiler to HNode            &      66.0 &             8.3 &           87.6 &             116.7 &          217.5 &               25.7 &                   63.1 \\
HNode to Building              &      84.7 &            23.4 &           78.1 &             123.3 &          265.8 &               57.4 &                   82.4 \\
HNode to Hot water             &      69.2 &            69.2 &           69.2 &              69.2 &           69.2 &               69.2 &                   69.2 \\
Pellet boiler to HNode         &       0.0 &             9.3 &            0.0 &               0.0 &            2.8 &                0.0 &                    0.0 \\
HNode to Heat storage          &      54.5 &            41.8 &           38.3 &              77.2 &           52.2 &               59.6 &                   52.9 \\
Heat storage to HNode          &      50.5 &            37.8 &           35.2 &              74.5 &           49.2 &               54.8 &                   49.3 \\
Photovoltaic to AC Node        &      89.7 &           102.4 &           94.7 &               0.0 &           90.4 &               98.9 &                   82.7 \\
Photovoltaic to FiTPV          &      24.1 &            33.9 &           26.4 &               0.0 &           19.2 &               27.3 &                   20.7 \\
Solar thermal to HNode         &       0.1 &             0.0 &            0.0 &               0.1 &            0.1 &                0.0 &                    0.0 \\
\bottomrule
\end{tabular}
}{Aggregated energy flows [TWh/a] between the different technologies for \textit{Value of} analysis. }{tab:value_of_flows}{Aggregated energy flows between the different technologies for Value of analysis}%

Table~\ref{tab:value_of_cap} shows the aggregated installed capacities for all residential buildings for the different sensitivity cases.

\myScaledTable{
\begin{tabular}{lrrrrrrr}
\toprule
  &  Min  &  No &  No &  No &  No &  No &  Forced \\
&  Cost &  Gas supply &  Fuel cell &  Photovoltaic &  Heat pump &  refurbishment &  refurbishment \\
\midrule
Gas boiler [GW$_{th}$]      &      36.6 &             5.1 &           47.6 &              83.9 &          111.9 &               14.7 &                   36.3 \\
Oil boiler [GW$_{th}$]      &       0.0 &             0.0 &            0.0 &               0.0 &            0.0 &                0.0 &                    0.0 \\
CHP        [GW$_{el}$]      &       0.0 &             0.0 &            0.0 &               0.0 &            0.0 &                0.0 &                    0.0 \\
District heating [GW$_{th}$]&       0.0 &             7.5 &            0.9 &               0.0 &           13.1 &                0.0 &                    0.0 \\
Heat storage     [GWh$_{th}$]&     215.6 &           291.4 &          216.6 &              63.4 &          173.9 &              265.4 &                  188.6 \\
Heat pump        [GW$_{th}$]&      60.4 &            83.1 &           60.4 &              47.0 &            0.0 &               95.5 &                   34.6 \\
Electric heater  [GW$_{th}$]&      99.7 &            97.1 &           90.3 &              83.6 &           75.4 &              121.0 &                   84.4 \\
Solar thermal    [GW$_{th}$]&       0.2 &             0.0 &            0.0 &               0.2 &            0.2 &                0.0 &                    0.0 \\
Pellet boiler    [GW$_{th}$]&       0.0 &             2.4 &            0.0 &               0.0 &            0.8 &                0.0 &                    0.0 \\
Battery          [GWh$_{el}$]&      16.9 &             7.1 &           12.8 &               0.0 &           17.1 &               20.0 &                   17.4 \\
Fuel cell        [GW$_{el}$]&      12.7 &             0.0 &            0.0 &              14.5 &            5.1 &               16.9 &                   10.7 \\
Photovoltaic     [GW$_{el}$]&     133.4 &           160.3 &          142.3 &               0.0 &          127.8 &              147.6 &                  121.0 \\
\bottomrule
\end{tabular}
}{Aggregated installed capacities of the different technologies for the \textit{Value of} analysis. }{tab:value_of_cap}{Aggregated installed capacities of the different technologies for the Value of analysis}%